\documentclass{article}

\usepackage{arxiv}

\usepackage[utf8]{inputenc} 
\usepackage[T1]{fontenc}    
\usepackage{hyperref}       
\usepackage{url}            
\usepackage{booktabs}       
\usepackage{amsfonts}       
\usepackage{nicefrac}       
\usepackage{microtype}      
\usepackage{lipsum}         
\usepackage{graphicx}
\usepackage{natbib}
\usepackage{doi}
\usepackage{enumitem}
\usepackage[english]{babel}

\usepackage{amsmath}
\usepackage{graphicx}
\usepackage{tikz}
\usepackage{comment}
\usepackage{wrapfig,booktabs}
\usepackage{cleveref}       

\usepackage{orcidlink}

\usetikzlibrary{arrows,shapes,positioning}
\usetikzlibrary{calc,decorations.markings}
\usetikzlibrary{decorations.pathreplacing}

\DeclareMathOperator*{\argmax}{argmax}

\title{On conceptualisation and an overview of learning path recommender systems in e-learning}
\usepackage{authblk}

\newbox{\orcid}\sbox{\orcid}{\includegraphics[scale=0.06]{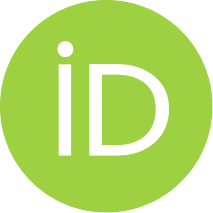}} 

\author[1]{%
	\href{https://orcid.org/0009-0003-3780-5064}{\usebox{\orcid}\hspace{1mm}A. Fuster-López}%
}
\author[1]{%
	\href{https://orcid.org/0000-0002-8847-8900}{\usebox{\orcid}\hspace{1mm}J.M. Cruz}%
}
\author[1]{%
	\href{https://orcid.org/0000-0003-3126-0078}{\usebox{\orcid}\hspace{1mm}P. Guerrero-García}%
}
\author[1]{%
	\href{https://orcid.org/0000-0003-1572-1436}{\usebox{\orcid}\hspace{1mm}E.M.T. Hendrix}%
}
\author[2]{%
	\href{https://orcid.org/0000-0001-6938-221X}{\usebox{\orcid}\hspace{1mm}A. Ko\v{s}ir}%
}
\author[3]{%
	\href{https://orcid.org/0000-0001-9527-3455}{\usebox{\orcid}\hspace{1mm}I. Nowak}%
}
\author[4]{%
	\href{https://orcid.org/0000-0002-8445-395X}{\usebox{\orcid}\hspace{1mm}L. Oneto}%
}
\author[5]{%
	\href{https://orcid.org/0000-0002-2997-4568}{\usebox{\orcid}\hspace{1mm}S. Sirmakessis}%
}
\author[6]{%
	\href{https://orcid.org/0000-0001-7915-0391}{\usebox{\orcid}\hspace{1mm}M.F. Pacheco}%
}
\author[6]{%
	\href{https://orcid.org/0000-0001-9542-4460}{\usebox{\orcid}\hspace{1mm}F.P. Fernandes}%
}
\author[6]{%
	\href{https://orcid.org/0000-0003-3803-2043}{\usebox{\orcid}\hspace{1mm}A.I. Pereira}%
}

\affil[1]{Universidad de Málaga, 29080 Málaga, \texttt{\{afuster,jmcruz,pguerrero,eligius\}@uma.es}}
\affil[2]{University of Ljubljana, 1000 Ljubljana, \texttt{andrej.kosir@fe.uni-lj.si}}
\affil[3]{Hamburg University of Applied Sciences, 20099 Hamburg, \texttt{ivo.nowak@haw-hamburg.de}}
\affil[4]{University of Genova, 16126 Genova, \texttt{luca.oneto@unige.it}}
\affil[5]{University of Peloponnese, 26334 Patras, \texttt{syrma@uop.gr}}
\affil[6]{Instituto Politécnico de Bragança, 5300-253 Bragança, \texttt{\{pacheco,fflor,apereira\}@ipb.pt}}

\renewcommand{\headeright}{Technical Report}
\renewcommand{\undertitle}{Technical Report}
\renewcommand{\shorttitle}{On Recommender Systems in E-Learning}

\hypersetup{
pdftitle={On Conceptualisation and an Overview of Learning Path Recommender Systems in E-Learning},
pdfsubject={cs.LG, cs.CL},
pdfauthor={A. Fuster-López, J.M. Cruz, P. Guerrero-García, E.M.T. Hendrix, A. Ko\v{s}ir, I. Nowak, L. Oneto, S. Sirmakessis, M.F. Pacheco, F.P. Fernandes, A.I. Pereira},
pdfkeywords={E-learning, recommender systems, learning paths, educational technology, conceptualisation},
}

\begin{document}
\maketitle

\begin{abstract}
The use of e-learning systems has a long tradition, where students can study online helped by a system. In this context, the use of recommender systems is relatively new. In our research project, we investigated various ways to create a recommender system. They all aim at facilitating the learning and understanding of a student. We present a common concept of the learning path and its learning indicators and embed 5 different recommenders in this context.

\end{abstract}


\section{Introduction}
In recent years, the landscape of e-learning has witnessed exceptional advancements, providing students with tools to improve their performance. In the pursuit of optimizing the e-learning experience, one emerging area of focus is the integration of recommender systems. By leveraging sophisticated algorithms, recommender systems aim to personalize the learning path by tailoring recommendations based on individual student performance, preferences, learning style and other factors. 

The iMath \cite{iMath} project aims at developing an  AI-driven tool to support the personalized learning path of students enrolled in higher education mathematical subjects. This tool 
should be capable of personalizing an e-learning path tailored to each student's unique needs, taking into account the current knowledge level and also ensuring a learning path that aligns with their individual requirements rather than solely relying on an expert perspective.

Throughout the iMath project, the authors investigated the question of how to personalise an e-learning path for an individual student. This path does not necessarily coincide with the one a teacher would choose, although this often is considered the best way to go in a teaching environment. The insights gained from the developed tools have the potential to be integrated in the MathE portal, cf.~\cite{MathE,pacheco2019mathe}, to enhance its educational aim.

This report describes insights from the iMath project with the collaborative efforts of various universities in this Erasmus+ project. Our collective goal is to establish a cohesive conceptual framework for learning path recommender systems. Through this comparative analysis, we aim not only at contributing to the evolving landscape of personalized e-learning, but also to offer valuable insights for the future development of recommender systems in educational contexts. The remainder of this paper is organized as follows. Section \ref{sec:formulation} describes the context our research question of an existing e-learning environment MathE. In Sect. \ref{sec:literature}, we embed the investigation into the literature on the topic. Section \ref{sec:methods} describes several recommenders that have been developed in the iMath project and Sect. \ref{sec:concl} summarizes our findings.

\section{Problem formulation}
\label{sec:formulation}

The integration of an effective learning path recommender system in an e-learning environment involves various aspects to be considered. Thus, we first focus on the conceptual and practical details associated with the core problem of personalizing a learning journey for an individual student.

\subsection{Conceptual problem formulation}

\begin{figure}[ht]
    \centering
    \includegraphics[page=1,width=\textwidth]{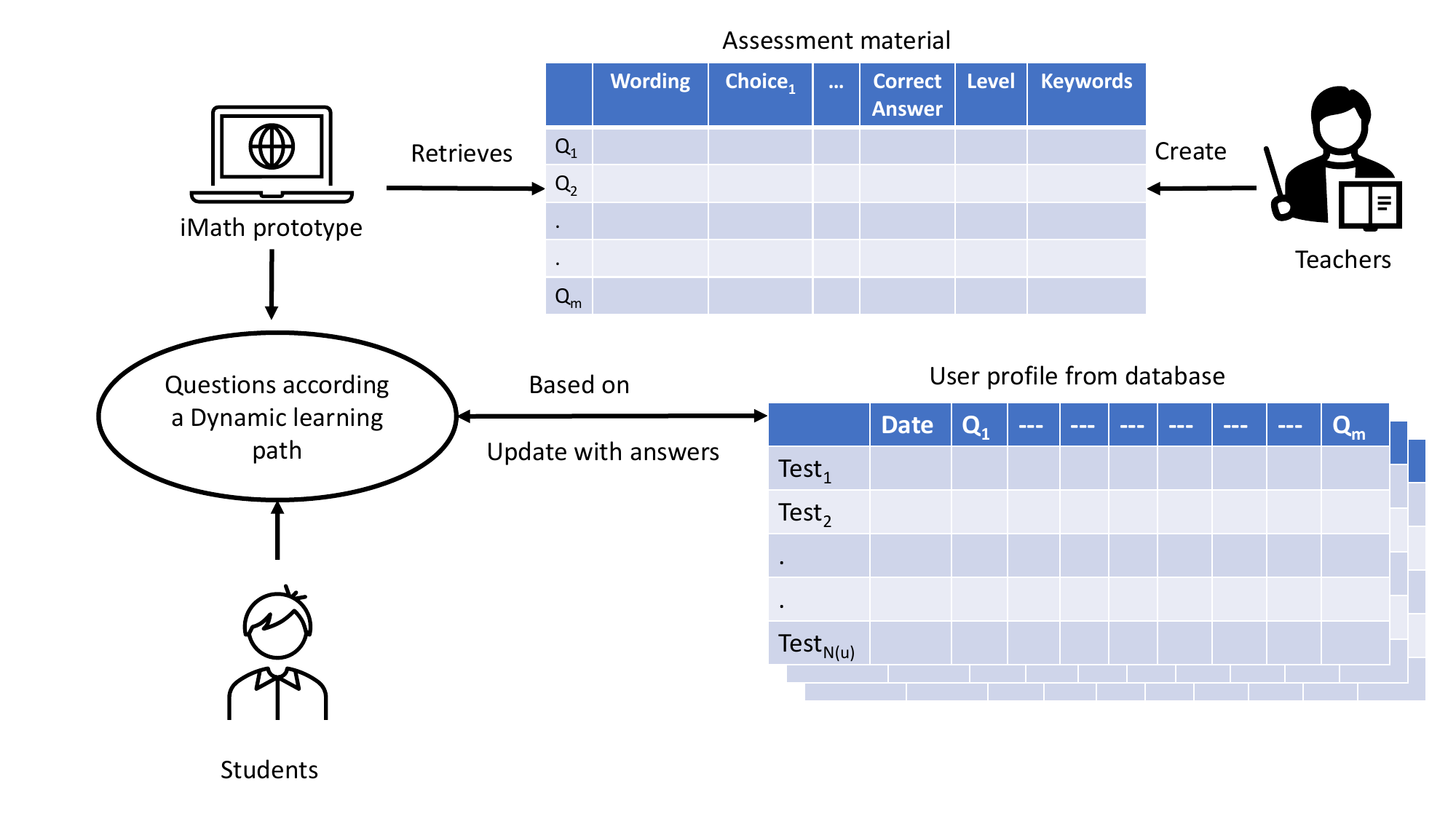}
    \caption{System overview}
    \label{fig:system-overview}
\end{figure}

Traditional methods, often relying on a predefined path shaped by expert perspectives, do not take the difference in learning style into account, i.e. preferences and knowledge level of individual students. Our main focus is to address the formulation of a recommender system that goes beyond static structures. Unlike other learning paths that may overlook the significance of assessment, our approach is integrated in the evaluation process seamlessly. Fig. \ref{fig:system-overview} depicts how users can access the iMath prototype, where they take multiple choice tests with each question chosen following one of the recommender methods presented in Section \ref{sec:methods}. 

By identifying the possible inputs and outputs of the systems with the available information, it is possible to apply different techniques such as the adaptation of a Keyword-Question matrix, analogous to a Term-Document matrix used in the field of Natural Language Processing \cite{indurkhya2010handbook} (NLP). By using such approach, it is possible to get to the essence of the learning indicators, representing the intersection between the educational content (in this case, keywords) and the assessment criteria (in this case, questions). The challenge then lies in obtaining an algorithm capable of providing the learner with a dynamic learning path, adjusted to the individual user interaction with the e-learning system. In this way, a recommendation is provided aligned with the user's unique learning profile. 

\subsection{Scope of the tool}

The MathE platform 
\cite{MathE} is an online educational system running since February 2019. Its objective is to assist students in Mathematics subjects taught in higher education institutions.
This platform is organized in 21 math topics/subtopics and provides free access to various resources, including videos, exercises, practice tests, and pedagogical material
\cite{mathEjournal}.

A well-known MathE tool is the online self-assessment test, where seven questions are selected on a given topic. 
How should these questions be selected in order for the student to maintain his motivation in math study? How should these questions be selected in order the student to learn a given topic? To attempt answering these questions, various methods outlined in section \ref{sec:methods} were implemented on top of a prototype mimicking the MathE self-assessment test, choosing five pertinent questions from a single topic.

The aforementioned prototype
contains two main student related data sources, educational content and assessment criteria. The first dataset primarily contains information regarding the questions that can be posed to students: the possible options for a multiple-choice test, the correct answer, the difficulty level and various identifiers for its keywords, enabling categorization and thematic analysis. The second dataset focuses on the tests realized by each student and contains information such as name, surname, university, and email, along with responses to a final satisfaction questionnaire of each test. Additionally, this dataset captures user interaction with the educational prototype, logging timestamps for button clicks and answers. Notably, the test responses and user interaction are aligned with specific question and answer indices from the former dataset. 

\subsection{Explanatory example}

To illustrate the problem formulation described above, consider a practical example from the point of view of the two main actors. In this e-learning environment, the students engage with the platform by taking multiple choice tests, each chosen according to one of the recommender methods described in Sect. \ref{sec:methods}.

\begin{itemize}[leftmargin=*]
    \item Student Perspective: Alice, a student, uses the iMath prototype to study linear algebra. Unlike traditional e-learning systems that offer a one-size-fits-all learning path, the proposed system dynamically adapts to Alice's performance and preferences. After completing a test with questions selected by one the developed recommender systems, Alice's interactions, such as answer choices and time spent on questions, are collected. This data is useful to tailor her future learning paths, making her study experience more personalized and effective.
    \item Educator Perspective: Bob, a teacher, contributes to the course by preparing multiple-choice questions, including possible answers, the correct answer, difficulty level, and relevant keywords. This information is crucial for the generation of the learning path, as it can be utilized by the recommender system.
\end{itemize}

In conclusion, this environment bridges the gap between students and educators, by leveraging advanced algorithms and user-generated data to offer a more personalized and engaging learning experience.

\section{Related work}
\label{sec:literature}

Given the relevance of both fields, the use 
of recommender systems in e-learning has gained significant attention. There are several notable reviews, such as \cite{klasnja2015recommender,tarus2018knowledge-based-ontology,zhang2022recommender} delving into various domains fairly related to the different approaches discussed in the following sections. 
The characterization of the recommendation techniques found in \cite{tarus2018knowledge-based-ontology} is quite useful to distinguish the different philosophies taken in each approach.


Group based recommender systems can harness the collective information from both users and items to generate recommendations. The system can analyze the learning material and categorize it, based on various features such as subject matter, difficulty level, learning style and many more. This approach can help in creating a structured learning path for the students, making the learning process more organized. Several frameworks have been devised; notably \cite{dwivedi2015grouprecom} in the domain of recommender systems utilized in e-learning environments. While our main focus is primarily on organizing learning material such as questions, the article shifts its attention to the categorization of students by utilizing neighbouring groups on user profiles.


Collaborative filtering is a widely used technique in recommender systems. This method generates recommendations based on the behaviour of similar users. This approach can be particularly effective in personalized learning, as it can help to identify resources that are likely to be of interest to the learner based on past behavior and the behavior of similar learners that might have faced the same problems while learning. \cite{bobadilla2009colfilter} aims to establish the foundation for applying this technique to an e-learning environment within the scope of memory-based approaches. Additionally, it examines potential metrics for measuring its performance. 


Utility based approaches resonate with the principles of dynamic optimization and adaptive decision-making. These systems provide recommendations based on the perceived value or utility they offer to individual users. This alignment fits the idea of continuously refining recommendations for maximizing the user satisfaction and learning outcomes. An ontology-based take can be found in \cite{zielinski2015utility} where the utility score is calculated for learning objects that form part of the learning pathway.


Content based techniques operate on the premise of assessing intrinsic attributes of learning materials and aligning them with the user preferences. This methodology delivers personalized content selection without explicitly asking for user feedback. By analyzing the characteristics and features of the learning materials, the system can create profiles for the content. 


The knowledge based approach incorporates domain knowledge into the recommendation process. By understanding the relationships between concepts within the learning domain, the system should be able to provide recommendations grounded in pedagogical principles.

\section{Developed methods}
\label{sec:methods}


Although the methods presented in this section are introduced in a self-contained way, giving the impression of being exclusive and differentiated from each other, it is worth noting that we can face the recommendation on two layers. An ``upper layer", closer to the learning indicators and the learning path, and a ``lower layer", more focused on working with the final recommendation to provide a higher level with tools to generate the learning pathway. Therefore, readers should note that these methods can also support each other to achieve a hybrid system.

\subsection{Concept map and graph walk-based method} 
\label{sec:concept}

A student activity recommendation system aims at recommending to students how they can best master a given content in terms of selected learning indicators in a given time.

\subsubsection{Basic approach}
The architecture of this recommendation system consists of two components. Firstly, a concept map is generated from the learning content as a directed graph, and secondly, the recommendation of student activities (in our particular case, questions) is a two-stage graph walk.

\begin{figure}[h]
    \centering
    \includegraphics[page=1,width=0.6\textwidth]{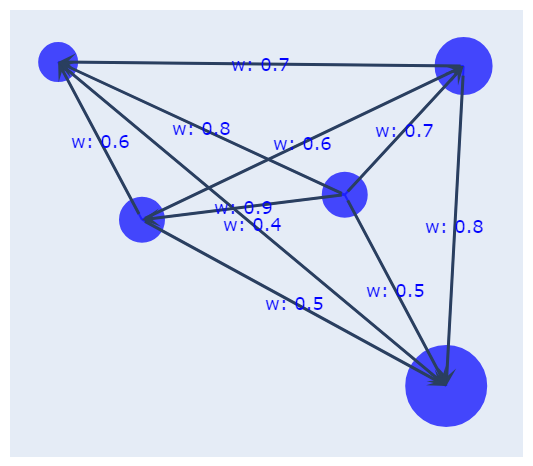}
    \caption{An example of the concept map. Arcs are directed and weighted. The size of the node is proportional to the number of recommended items this concept involves.}
    \label{fig:concept_map_example}
\end{figure}

The concept map is defined as a weighted directed graph. The nodes are concepts labeled by a set of content pieces on which students should work (in our case individual questions). Directed arcs indicate which concept a student should master first. Arc weights indicate how strong the dependence of these two nodes is. An example of a concept is given at Fig. \ref{fig:concept_map_example}. 

The student activity recommender system prescribes content pieces called recommender items by which concept map nodes are labeled. Each recommendation can be seen as a two-level graph walk where the last visited item is recommended. At the top level, the walk is a tour among concepts. This walk proceeds from this to the next concept when this concept is mastered by the student. Explicit criteria for mastering a concept are given such as the threshold of the proportion of correctly answered questions etc. When a given concept is first entered, a bottom-level walk is done inside the concept. Selected one or more learning indicators are estimated for each of the candidate recommendation items and the next recommended item is the one having the best profile of learning indicators. 

There are two main challenges to this approach. The first is an automatic generation of concept maps from learning content (textual descriptions, questions, keywords etc) using machine learning algorithms, see \cite{li2019concept}. Real content testing shows that automatically generated concept maps usually need manual correction to remove some obvious anomalies. However, automatically generated concept maps may be combined with human-generated ones. In fact, a teacher may manually generate an adequate concept map. Obviously, this approach is highly dependent on the quality of the underlying concept map. 

The next challenge is an automatic estimation of student learning indicators. Some correlates to learning indicators are easy to estimate such as the proportion of correct answers, the coverage of the concept, the probability of the correct answer to a given question etc. Methods based on clickstream \cite{liu2022predicting} may provide good results. 

To summarize this recommendation algorithm as a graph walk, the top level is a variant of a modified depth-first walk selecting an optimal path through the concepts to promote an adequate order of concept mastering. The bottom level is a greedy approach where the next recommendation item is the one for which the estimated or expected learning outcome profile is the best.

The selected approach to the recommender system facilitates the following parametrizations:
\begin{itemize}[leftmargin=*]
  \item The profile of explicitly maximized learning outcome indicators are selected arbitrarily to configure the properties of the recommender.
  \item Any concept map can be used, machine or human-generated, or a combination of these ones. 
  \item Criteria to measure whether a given concept is mastered and when a given content (full concept map) is mastered is also selected arbitrarily.
  \item Personalization and contextualisation of the recommender system can be integrated naturally into the recommender algorithm at two stages. First, a concept map may be generated for an individual student, and second, learning outcome indicators may be estimated in a personalized and contextualized way.
\end{itemize}

\subsubsection{Input and output}
The input to the algorithm is learning content such as set of questions, keywords, textual description of concepts etc. Modern machine learning techniques for content-based image retrieval and generative model-based image description could utilize visual content as well.

The output of the approach is a sequence of recommendation items, that is a sequence of content pieces a student should work on to master the content in a short amount of time. 

\subsubsection{Advantages and drawbacks}
The advantages of this approach are the following: 
\begin{itemize}[leftmargin=*]
  \item Selected learning outcome indicators are maximized directly.
  \item Human (teacher) view of course content structure can be combined with the machine-generated concept maps. On the other hand, the recommender system can be established independently from the teacher.
  \item Criteria for mastering a concept and mastering the whole concept can be selected explicitly.
  \item Latest advance approaches to concept map generation (generative models) and to automatic estimation of learning outcomes indicators can be integrated.
  \item Personalisation and recommendation of recommender system is straightforward.
\end{itemize}

The main drawback of the approach is the sensitivity to the quality of the concept map. Clearly, an inadequate concept map will produce an ineffective sequence of actions. The remedy is a combination of machine and human-generated concept maps.

\subsection{Collaborative filtering recommender} 

This recommendation system aims at providing questions based on their performance when completing the tests. 

\subsubsection{Basic Approach}

Collaborative filtering (CF) plays with the assumption that users who have similar preferences in the past are expected to have similar preferences in the future. In the context of e-learning we can think about these ``preferences" as the current situation of the learner in a learning path, where a student might face the same challenges, lack of basis in a previous concept or struggles as another student. 


The objective of these CF systems is to minimize the prediction errors \cite{bobadilla2009colfilter}. A handy implementation of the methodology is provided by the Python library Surprise \cite{hug2020surprise}. The library provides several recommender system algorithms. Some of them are tailored for collaborative filtering by providing prediction algorithms and the ability to build your own. These prediction algorithms take the historical data on user-question interaction as input. A user-question relationship is organized in a matrix similar to the one shown in Table \ref{table:user-question}.

\begin{wraptable}{L}{5.5cm}
\caption{User-question matrix}
\label{table:user-question}
\centering
\begin{tabular}{|l|c|c|c|c|} 
\hline
        & $\text{Q}_1$ & $\text{Q}_2$ & … & $\text{Q}_m$  \\ 
\hline
$\text{User}_1$ & 3    & 4    &   &       \\ 
\hline
$\text{User}_2$ &      & 2    &   & 5     \\ 
\hline
$\text{User}_3$ &      & 4    &   & 3     \\ 
\hline
\multicolumn{5}{|c|}{. . .}       \\ 
\hline
$\text{User}_u$ & 1    & 2    &   &       \\
\hline
\end{tabular}
\end{wraptable}


This matrix holds the user rating data for each question, which is then used to calculate the similarity between users. The similarity can be calculated using various techniques, which  can be memory-based or model-based. Memory-based approaches directly compare users or items based on their past interaction by using similarity metrics in algorithms like k-nearest neighbors (KNN). Predictions for a user on items they have not interacted with are made by aggregating the ratings or preferences of similar users or items. Moreover, model-based techniques use algorithms like singular value decomposition (SVD) or probabilistic models to learn patterns and relationships from the data.
%
%
In this implementation, we used a hybrid approach computing similarity among users while combining both memory-based and model-based collaborative filtering techniques. Specifically, the system employed singular value decomposition (SVD) and k-nearest neighbors (KNN) to calculate similarities among users based on their question-solving behavior. This hybridization aims at leveraging the strengths of both approaches, which can potentially improve recommendation accuracy.


One of the primary considerations during the design of the rating matrix was determining the most appropriate method for users to ``rate" a question. An explicit method would be using the opinion provided by students for the questions they answered in tests like they would a film or song. Such an approach is impractical. Thus, an implicit feedback approach was chosen considering data from the assessment material, such as the difficulty of the questions and whether the student response was correct or incorrect. This rating system utilizes an integer valuation ranging from 1 to 5: 


\begin{itemize}
    \item Value 1 indicates that the student answered with ``I don't know".
    \item Values 2 and 3 represent wrong and right answers, respectively, for a basic question,
    \item Values 4 and 5 represent wrong and right answers, respectively, for a difficult question.
\end{itemize}

\noindent By employing this implicit feedback mechanism, the system evaluates the current state of knowledge regarding a specific set of questions. It adopts a more lenient approach towards students providing incorrect answers to difficult questions, while penalizing those who give wrong answers to basic questions. In another way, the system would reward correct responses to difficult questions without overemphasizing the significance of answering a basic question correctly.

\begin{figure}
  \centering
  \resizebox{0.5\textwidth}{!}{%
  \begin{tikzpicture}[node distance=1.5cm and 1.5cm, mainconcept/.style={circle, draw, minimum size=2cm}]
    \node[mainconcept] (A) {Matrices};
    \node[below left =0.25cm and 0.15 cm of A.north] (Q1) {$Q_{17}$};
    \node[above right=0.25cm and 0.10 cm of A.south] (Q2) {$Q_{31}$};
        
    \node[above right=1 and 0.5 cm of A, mainconcept] (B) {Determinant};
    \node[above right= 0.25cm of B.south] (Q7det) {$Q_7$};
    \node[below = 0.25cm of B.north] (Q5) {$Q_5$};

    \node[right=2 cm of A, mainconcept] (C) {Eigenvalue};
    \node[above= 0.25cm of C.south] (Q7eigen) {$Q_7$};

    \node[right=2 cm of B, mainconcept] (D) {Multiply};
    \node[below right=0.2 and 0.15 cm of D.north] (Q11) {$Q_{11}$};
    \node[below left=0.2 and 0.15 cm of D.north] (Q14) {$Q_{14}$};

    \draw[->] (A.north) to (B.west);
    \draw[->] (B.south west) to (A.north east);

    \draw[->] (A) to (C);

    \draw[->] (B) to (C);

    \draw[->] (B) to (D);
    \draw[->] (C) to (D);

    \node[align=center,below= 0.5 cm of C] (Text) {Find optimal question\\ tailored  to a specific\\ concept and student};

    \draw[->] (Text.north east) to (Text.south east);
    \draw[->] (Text.south west) to (Text.north west);

    \node[draw, rectangle, minimum size= 3cm,below= 1 cm of Text] (CF) {CF Reccomender};

  \end{tikzpicture}
  }%
    \caption{System interoperability}
    \label{fig:system-interoperability}
\end{figure}
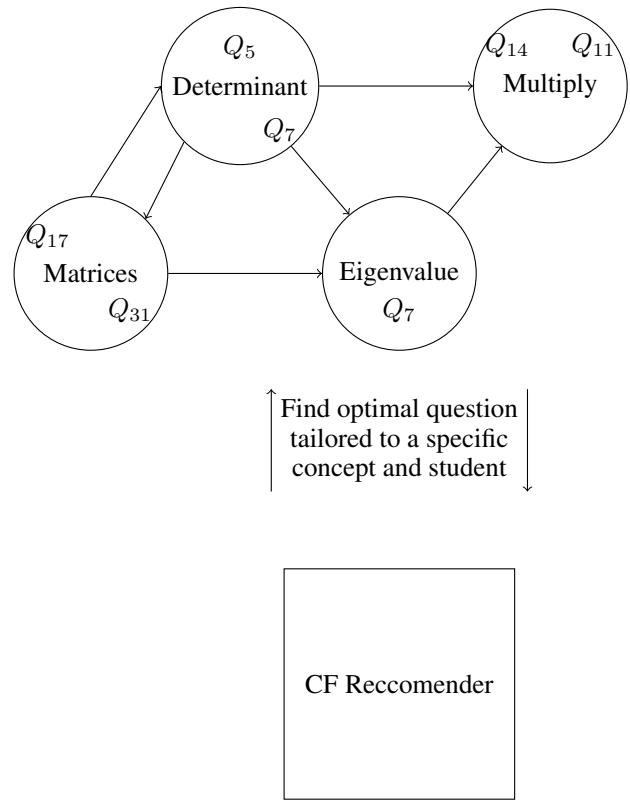

While this approach can function independently by providing students with questions based on their peers' performance at similar knowledge levels, its true strength lies in its modularity. Since it relies heavily on the assessment material provided by the teacher, it allows flexibility and customization according to specific educational needs and goals. This adaptability can be further enhanced by an upper layer of a fully-fledged system, enabling the integration of additional features and functionalities tailored to the educational context. For instance, an upper layer focused on optimally traversing a concept graph could use this approach to find the best question for a given concept based on the current state of the student, as depicted in Fig. \ref{fig:system-interoperability}. This lower layer is also compatible with other approaches, such as a personalized learning path recommendation system that utilizes machine learning algorithms to further analyze individual learning patterns and improve the learning path.


\subsubsection{Input and output} 

The system retrieves all previous responses provided by a student and matches them with the information contained in the assessment material. It then generates formatted data, comprising a table with rows containing user ID, question ID, and the evaluation of the response, which serves as input of the collaborative filtering algorithm.

The algorithm provides a list of tuples (question ID, estimated rating for prediction) and sorts them based on the highest rating.

\subsubsection{Advantages and drawbacks}


The advantages of the collaborative filtering approach for question recommendation include:

\begin{itemize}
    \item Personalized recommendations based on past behavior and current status of knowledge.
    \item Adaptability to evolving user preferences over time.
    \item Potential for augmentation through the integration of additional layers focused on alternative learning metrics or material insights.
\end{itemize}

The main drawbacks of this approach are:

\begin{itemize}
    \item Cold start problem for new users or questions.
    \item Difficulty in capturing complex user behavior by itself.
    \item Dependency on a substantial dataset of questions to optimize performance.
\end{itemize}

\subsection{Clustering-based method} 

The method is based on clustering techniques combined with recommender systems using graph theory.

\subsubsection{Basic approach}

In this approach, the $k$-means clustering algorithm was used to classify the questions according to their difficulty level. Clustering is an unsupervised data partitioning method aiming at the classification of elements of a dataset into groups (clusters) based on similarity and dissimilarity among features of the elements. In this way, it aims at discovering underlying patterns in an unsupervised manner~\cite{rehman2022divide}.


Graph theory is utilized to model the relations among the different learning elements within the platform. In this model, questions and keywords are represented as nodes and the arcs indicate the associations between them. 

The process is based on four phases. In Phase I, the score of the question is 
determined by combining feedback from teachers and students. After that, 
Phase II employs the $k$-means clustering algorithm to categorize questions into distinct difficulty levels.
In Phase III, 
the algorithm constructs a graph to map the connections between questions and keywords, where the degree of each node is helpful to identify the most relevant successive questions for the learner. 
Finally, phase IV selects 
the most appropriate next question.

\subsubsection{Input and output}

The input data includes questions and keywords, the level of each question given by the teacher, and the number of attempts and correct answers for each question. 
The output is the next question in the self-assessment test.

\subsubsection{Advantages and drawbacks}

The advantages of the presented algorithm include:
\begin{itemize}

\item Easy to implement.

\item Compromise between teacher and student feedback.

\item Possibility to explore ways to find the perfect personalized learning path for each topic/subtopic.

\end{itemize}

The main drawbacks of the presented algorithm are:
\begin{itemize}

\item Requires a sufficient distribution of keywords among the questions.

\item High probability of obtaining cluster levels without questions not yet answered.

\end{itemize}

\subsection{Supervised learning-based method} 
\label{sec:SL}

This approach explores the use of supervised machine learning in combination with a simple recommender system.

\subsubsection{Basic approach}

The approach selects a given question as follows:
\begin{itemize}
\item First question: based on the available background information about the students we train a Random Forest~\cite{breiman2001random} to predict, for each possible question, the probability of the student to answer the question correctly and the time needed to answer it.
Then, based on an simple heuristic rule, the next question is selected such that it gives the best compromise between the probability to answer correctly and time spent on the question.
\item From the second to the last question: we exploit the same procedure of the first question enriching the features on which we train the Random Forest with the information collected during the previous questions.
\end{itemize}

The idea behind this approach is to increase the motivation of the student in studying the subject, promoting staying longer on the platform, presenting questions with a high probability to be answered correctly.
In this way, the students, repeating the self assessment test, will receive first the simplest questions and then, with time, the more difficult ones since the simplest (and already provided to students) are removed from the set of possible choices.
\subsubsection{Input and output}
The data exploited by Random Forest is coming from two main sources:
\begin{itemize}[leftmargin=*]
\item Background data from the informative questionnaire filled in by each student.
User grades have been normalized to a scale between 0 and 100 to make grades of different scale comparable. In case of not answered questions or in case the student did not take any math exams yet, the missing data have been filled up with the mode of all the other user answers. 
\item Data from previous questions and tests on the platform.
For each past question of the current test, the system collects: if the answer is correct or not, if it has been skipped or not, the time spent on the question, difficulty of the question indicated in the input data, the percentage of times the users correctly answered the question, number of clicks on the answers.
\end{itemize}

As output, the algorithm gives the probability of answering the question correctly  and the required time to answer.
\subsubsection{Advantages and drawbacks}
The advantages of the proposed algorithm include:
\begin{itemize}
\item The approach is purely data-driven and does not require any additional side information.
\item The approach is simple to implement and can scale well to the addition of new feature or information.
\item The system can be updated easily and periodically as soon as the platform collects more data of new users.
\item The approach is computationally cheap to use in real time since predictions are actually computationally inexpensive.
\end{itemize}
The main drawbacks of the proposed algorithm are:
\begin{itemize}
\item The fact that it is purely data-driven may lead to unpredictable results.
\item The heuristic rule to recommend the next question is too simplistic.
\item The difficulty to consider side information.
\end{itemize}
\subsection{Reinforcement learning-based method} 

Reinforcement learning (RL) offers a framework for learning optimal policies in uncertain environments. When applied to an e-learning environment, RL treats the problem as a Markov Decision Process (MDP), where the system learns to make decisions based on past experiences 
\cite{fahad_mon_reinforcement_2023,paduraru_using_2022,pu_deep_2020,tang_systematic_2024}. 
The goal is to maximize the expected total learning achievement of a student's learning path over a certain number of time steps.
In order to compute an optimal learning path,
RL computes a predictive model, which helps to find the best learning path, considering path constraints, like learning all concepts in a given number of time-steps.

\subsubsection{Basic Approach}
In this approach, the learning environment is modeled as an MDP consisting of:
\begin{itemize}[leftmargin=*]
\item States: Represent the context of the system at a certain time step. The context is composed of the learner's knowledge status, preferences, past interaction, and relevant information of the system, including data from other students.
\item Actions: Correspond to the recommendation of the learning material at each time step, such as the question asked in that time step.
\item Environment: Dictates the outcome of a state-action pair by providing a new state
following a so-called transition function, which may include uncertainty.
\item Rewards: Indicate the feedback received by the learner after taking one or several actions. A dense reward
provides an immediate feedback after each action, such as the correctness of the answer, time taken to respond, the difficulty level, or any value for selecting
a question, as described in Sections
\ref{sec:concept}-\ref{sec:SL}.  A sparse reward is
computed after several actions or after finishing
a complete learning path, e.g., by checking if the path includes all required learning concepts.
\end{itemize}
RL computes a predictive model defined by so-called $Q$-values for each state-action pair, which are used to select an action at a given state by maximizing the $Q$-value. This defines a policy that determines the sequence of learning materials to present to the student.
Since the system data is dynamically changing,
 the $Q$-values are updated in each time step.
The approach can
be considered as an operational planning method of learning material
with the goal always to be sure, that the operational plan provides a feasible  learning path.

\subsubsection{Input and Output}
The input to the RL agent includes learner data such as past test performances, learning preferences, context information, and the characteristics of the available learning materials, similar as in
Sections \ref{sec:concept}-\ref{sec:SL}. 
The output of RL are $Q$-values for each state-action pair, which are used to select a learning action.

\subsubsection{Preliminary Experiment} In order to test the performance of a simple RL-algorithm for question selection,  a Dijkstra shortest path algorithm was adapted for selecting questions covering
12 topics of linear algebra. A concept map was defined for specifying  a partial ordering of topics using
an Excel table. A dense reward was computed for each state-action pair by providing the relative number of failures per question. A sparse reward was computed after finishing a complete learning path, by checking if the path includes all required learning concepts. 
The experiments have shown that the  adjustments of the Dijkstra shortest path algorithm  did not significantly increase the computational effort.

\subsubsection{Advantages and Drawbacks}
Some advantages of this approach include:
\begin{itemize}
\item Systematic approach to explore a huge set of learning paths by computing rewards for state-action pairs, as in
Sections \ref{sec:concept}-\ref{sec:SL}, as well as rewards
for complete learning paths.
\item In contrast to greedy approaches,
the policy computed by $Q$-values
provides feasible learning paths.
\item Personalized learning paths tailored to individual learner characteristics and preferences.
\item Adaptability to changes in the learner's knowledge level and learning trajectory while also handling learning path constraints, such as covering all concepts in order.
\item Maximization of learning outcomes and engagement.
\end{itemize}
The drawbacks of this approach are:
\begin{itemize}
\item Implementing RL requires expertise in algorithm design, training, and optimization.
\item For computing an effective reward function, methods
like in Sections
\ref{sec:concept}-\ref{sec:SL} are required, which need a substantial amount of data.
\item RL models can be opaque, which may make it difficult for teachers or students to interpret the reasoning behind the system recommendations.
\end{itemize}

\section{Summary and Conclusion}
\label{sec:concl}


The development of recommender systems in e-learning is an interesting challenge. In this paper, we described several approaches which have been tested on the same group of students.  Although the gathered data in the system is the same for all methods, the used information differs and the recommendation is based on varying underlying methodologies. We distinguish the focus on an upper layer related to learning outcomes and a lower layer focussing on generating a next (set of) questions or suggestion for learning material. The methods are diverse, each with its own approach to personalizing the learning experience for each student. Table \ref{tab:summary} summarizes their features to provide a clear overview.

\begin{table}[ht]
\centering
\caption{Summary}
\label{tab:summary}
\begin{tabular}{p{3cm}|p{3cm}|p{3cm}|c}
\textbf{Method}                  & \textbf{Input}                                                                  & \textbf{Output}                                                   & \textbf{Layer} \\ \hline
Concept Map and Graph Walk-Based & Learning content                                    & Sequence of items                                     & Top            \\ \hline
Collaborative Filtering          & User-question data                                                  & List of questions                      & Bottom         \\ \hline
Clustering-Based                 & Questions, keywords, difficulty levels, student performance data                & Next question in the self-assessment test                         & Top            \\ \hline
Supervised Learning-Based        & Background data from students; data from previous questions/tests               & Probability of answering correctly and required time to answer    & Bottom         \\ \hline
Reinforcement Learning-Based     & Learner data, learning material characteristics & $Q$-values to select learning actions & Top            \\ 
\end{tabular}
\end{table}

A significant insight from this research is the potential synergy between top-layer and bottom-layer systems. Top-layer systems (such as the ones pointed out in Table \ref{tab:summary}) focus on optimizing the learning path by considering the conceptual landscape of a subject by relying on keywords and information that could be difficult to be agreed among a group of teachers. These systems ensure that students engage with topics while maximizing understanding and retention. 

In contrast, bottom-layer systems generate specific recommendations based on more granular and easily measurable data, such as past performance or student preferences. These systems are powerful tools for personalizing the learning experience at a micro-level, ensuring that a single question is optimally suited to the student's current status. 

The potential for innovation lies in the integration of these two layers. By combining the big-picture focus of the top-layer systems with the detailed data-driven bottom-layer systems, it is possible to create an effective e-learning environment. This integrated approach allows for the dynamic adjustment of learning paths based on real-time feedback and performance data, ensuring that the learning experience remains engaging, challenging and tailored to the individual needs of each student.

In future studies, we will focus on the outcomes of the used recommenders for the same group of students. This requires a systematic design of experiments and measurable performance indicators. 

\bibliographystyle{unsrtnat}
\bibliography{recomender}

\end{document}